\documentclass{article}
\usepackage{spconf,amsmath,graphicx}

\usepackage{bbm}
\usepackage{caption}
\usepackage{enumitem}
\usepackage{tikz}
\usetikzlibrary{positioning}
\tikzstyle{block} = [inner sep=5pt, rectangle, draw, rounded corners, minimum height=20pt, text depth=0pt, text centered]

\newcommand{\indfun}[1]{\ensuremath{\mathbf{1}_{\{#1\}}}}

\title{Automatic Conflict Detection in Police Body-Worn Audio}

\name{Alistair Letcher$\,^{1, \dagger}$ \qquad Jelena Tri\v{s}ovi\'{c}$\,^{2, \dagger}$ \qquad  Collin Cademartori$\,^3$ \qquad Xi Chen$\,^4$ \qquad Jason Xu$\,^5$ \thanks{This research was supported by the LAPD and the UCLA Institute for Pure and Applied Mathematics, NSF Grant DMS-0931852.}}

\address{$^\dagger$ Joint first authors \quad $^1\,$University of Oxford \quad $^2\,$University of Belgrade \\ $^3\,$Brown University \quad $^4\,$Carleton College \quad $^5\,$University of California, Los Angeles}

\begin{document}
%
\maketitle
\begin{abstract}
Automatic conflict detection has grown in relevance with the advent of body-worn technology, but existing metrics such as turn-taking and overlap are poor indicators of conflict in police-public interactions. Moreover, standard techniques to compute them fall short when applied to such diversified and noisy contexts. We develop a pipeline catered to this task combining adaptive noise removal, non-speech filtering and new measures of conflict based on the repetition and intensity of phrases in speech. We demonstrate the effectiveness of our approach on body-worn audio data collected by the Los Angeles Police Department.
\end{abstract}
\section{Introduction}
\label{sec:intro}

Body-worn technology is beginning to play a crucial role in providing evidence for the actions of police officers and the public \cite{Ariel}, but the quantity of data generated is far too large for manual review.
In this paper we propose a novel method for automatic conflict detection in police body-worn audio (BWA). Methodologies from statistics, signal processing and machine learning play a burgeoning role in criminology and predictive policing \cite{mohler2015}, but such tools have not yet been explored for conflict detection in body-worn recordings. Moreover, we find that existing approaches are ineffective when applied to these data off-the-shelf.

Notable papers on conflict escalation investigate speech overlap (interruption) and conversational turn-taking as indicators of conflict in \textit{political debates}. In \cite{Grezes}, overlap statistics directly present in a hand-labelled dataset are used to predict conflict, while \cite{Caraty} detect overlap through a Support Vector Machine (SVM) with acoustic and prosodic features. The work in \cite{Kim} compares variations on both methods. Using automatic overlap detection, their method achieves $62.3 \%$ unweighted conflict accuracy at best in political debate audio. This approach is all the less effective on BWA data, which is far noisier and more diverse. Besides being harder to detect, overlap serves as an unreliable proxy for conflicts between police and public: these often involve little to no interruption, especially in scenarios where the officer is shouting or otherwise dominating the interaction.

We propose new metrics that successfully predict conflict in BWA along with speech processing and modeling techniques tailored to the characteristics of this data.
Section \ref{sec:pre} details adaptive pre-processing stages, feature extraction, and a SVM for non-speech discrimination.
In Section \ref{sec:repetition}, we develop metrics based on repetition using audio fingerprinting and auto-correlation techniques. This is based on the observation that conflict largely occurs in situations of non-compliance, where the officer repeats instructions loudly and clearly.
Finally, performance is evaluated on a dataset of $105$ BWA files provided by the Los Angeles Police Department (LAPD) in Section 5. The illustration below summarizes our conflict detection procedure.

\begin{figure}[h]
\begin{tikzpicture}
	\node[block] (1) at (0,2) {Denoising};
	\node[block] (2) [right=0.4cm of 1] {Feature Extraction};
	\node[block] (3) [right=0.4cm of 2] {Non-Speech Filter};
	\node[block] (4) at (2,0.5) {Repetition Detection};x
	\node[block] (5) [right=0.8cm of 4] {Conflict Score};

	\draw [->] (1) edge (2) (2) edge (3) (4) edge (5);
	\draw[->, rounded corners] (3.south) |- +(-6.3cm, -0.4cm) |- (4.west);
\end{tikzpicture}
\end{figure}

\section{Pre-Processing and Filtering}
\label{sec:pre}

The success of our approach relies on pre-processing steps catered to the task at hand. We apply adaptive denoising procedures followed by feature extraction for supervised discrimination of non-speech, also called Voice Activity Detection.

\subsection{Denoising}
\label{ssec:denoising}
Persistent noise like traffic, wind and babble as well as short-term bursts of noise including sirens, closing doors and police radio are present along with speech in BWA audio.
We filter persistent but non-stationary background noise based on optimally-modified log-spectral amplitude (OM-LSA) speech estimation, and apply minima controlled recursive averaging (MCRA) as described in \cite{Cohen}. Briefly, this approach computes the spectral gain while accounting for speech presence uncertainty, ensuring that noise removal best preserves speech components even when the signal-to-noise ratio (SNR) is low.

Let $x(n)$ and $d(n)$ denote speech and (uncorrelated, additive) noise signals respectively. Then the observed signal is $y(n) = x(n) + d(n)$, where $n$ is a discrete-time index. 
The spectrum is obtained by windowing $y(n)$ and applying the short-term Fourier transform (STFT), denoted $Y(k,l)$ with frequency bin $k$ and time frame $l$.
The STFT of clean speech $X(k,l)$ can be estimated as $\hat{X}(k,l)=G(k,l)Y(k,l)$, where $G(k,l)$ is the spectral gain function. Via the LSA estimator, we apply the spectral gain function which minimizes
\[ E\left[(log|X(k,l)|-log|\hat{X}(k,l)|)^2\right] \, . \]
Let hypotheses $H_0(k,l)$ and $H_1(k,l)$ respectively indicate speech absence and presence in the $k$th frequency bin of the $l$th frame. Assuming independent spectral components and STFT coefficients to be complex Gaussian variates, the spectral gain for the optimally modified LSA is given by
\[ G(k,l)={G_{H_1}(k,l)}^{p(k,l)}G_{min}^{1-p(k,l)} \,. \] 
Here $G_{H_1}(k,l)$ represents the spectral gain which should be applied in the case of speech presence and $G_{min}$ is the lower threshold for the gain in case of speech absence, preserving noise naturalness. $p(k,l)$ is the \textit{a posteriori} speech probability $P(H_1(k,l)|Y(k,l))$, computed using the estimates of noise and speech variance $\lambda_d(k,l)$ and $\lambda_x(k,l)$, the \textit{a priori} SNR $\xi(k,l)=\frac{\lambda_x(k,l)}{\lambda_d(k,l)}$ and the \textit{a priori} speech absence probability $q(k,l)=P(H_0(k,l))$. 

To estimate the time-varying spectrum of non-stationary noise $\lambda_d(k,l)$, we employ temporal recursive smoothing during periods of speech absence using a time-varying smoothing parameter. The smoothing parameter depends on the estimate of the speech presence probability, obtained from its previous values and the ratio between the local energy of the noisy signal and its derived minimum. Given $\hat{\lambda}_d(k,l)$ we may immediately estimate the \textit{a posteriori} SNR, 
\[ \gamma(k,l)=\frac{|Y(k,l)|^2}{\lambda_d(k,l)} \,. \]
This is used to estimate the \textit{a priori} SNR given by
\begin{align*}
\hat{\xi}(k,l) &= \alpha {G_{H_1}(k,l-1)}^2\gamma(k,l-1) \\
& \quad +(1-\alpha)\max\left\{\gamma(k,l)-1,0\right\} ,
\end{align*} 
with weight $\alpha$ controlling the noise reduction and signal distortion. The estimate $\hat{\xi}(k,l)$ allows for computing the probability of \textit{a priori} speech absence as described in \cite{Cohen}, which finally enables computation of the spectral gain and in turn speech spectrum.

We perform this filtering method three times in sequence to reliably remove residual noise that may persist after one stage of filtering. Doing so produces excellent results, eliminating most persistent noise while crucially avoiding attenuation of weak speech components. Nevertheless, sudden bursts of noise are rarely eliminated because the filter cannot adapt in time. We apply the method below to remove them, which is equally crucial to reliable repetition detection.

\subsection{Feature Extraction and Non-Speech Filter}
\label{ssec:feature}
The task of this section is to filter remaining non-speech. To begin, the audio signal is split into overlapping frames of size $0.06$s and with $0.02$s steps between start times. Over each frame, we compute $23$ short-term features consisting of the first $13$ Mel-Frequency Cepstral Coefficients (see \cite{Prahallad}); zero-crossing rate; energy and energy entropy; spectral centroid, spread, entropy, flux and rolloff; fundamental frequency and harmonic ratio. Features which require taking the Discrete Fourier Transform are first re-weighted by the Hamming window. Since many meaningful speech characteristics occur in a longer time-scale, we additionally include the mid-term features obtained by averaging our short-term features across frames of size $0.3$s and step $0.1$s.

We apply a SVM with Radial Basis Function kernel \cite[Chap. 12]{Hastie} to discriminate between speech and non-speech in this feature space. The SVM is trained on $38$ minutes ($22733$ frames) of labeled speech and $47$ minutes ($28239$ frames) of non-speech from BWA data. 
To evaluate predictive power we perform cross-validation (CV) with $10$ folds \cite[Chap. 7]{Hastie}. Our results are displayed in Table \ref{table:nonspeech} and compare favourably with state-of-the-art papers in speech detection, which obtain error rates no lower than $5\%$ in \cite{Elizalde} and $12\%$ in \cite{Zou}, on clean and noisy data (SNR at least 15 dB) respectively. Their learning algorithms include SVMs, Gaussian Mixture Models and Neural Networks.

\setlength{\tabcolsep}{13pt}
\renewcommand{\arraystretch}{1.5}
\begin{table}[htbp]
\centering
\begin{tabular}{ c | c | c  }
False Positive & False Negative & Total Error\\ \hline
$1.26\%$ & $3.61\%$  & $2.31\%$\\
\end{tabular} \caption{$10$-fold CV error in speech/non-speech detection.} \label{table:nonspeech} \vspace{-10pt}
\end{table}

\section{Repetition Detection and Scoring}
\label{sec:repetition}

Having eliminated most of the non-speech and noise, we turn to detecting repetitions as a measure of conflict. We split the audio into regions of interest and compare them using fingerprint and correlation methods based on \cite{Haitsma} and \cite{Herley}.

\subsection{Segmentation and Repetition}
\label{sec:segmentation}
In order to reduce the time it takes to search for repetitions, we automatically break the signal into regions which contain entire syllables, words, or phrases. We begin by applying a band-pass filter between $300$ and $3000$Hz, which we found to carry the most information about speech in our recordings.

Let $E'(t)$ be the energy (squared amplitude) of the signal in a window of length $0.05$s starting at time $t$, and define $E(t) = \indfun{E'(t) > 0.05} E'(t)$. This threshold filters windows with energy below $0.05$, in which the signal-to-noise ratio is too small for reliable repetition detection. We define points $t_0 = 0$ and $t_1,\ldots,t_n$ by the following criteria:
\begin{enumerate}[itemsep = 1pt]
\item $E(t_i)$ is a local minimum of $E(t)$.
\item There exists $t \in [t_{i-1},  t_i]$ such that $E(t) \neq E(t_i)$.
\item Each $t_i$ is the earliest time satisfying $(1)$ and $(2)$.
\end{enumerate}
\noindent This somewhat cumbersome definition deals with the possibility that $E$ attains the same local minimum value at consecutive times, by taking the earliest such time. We define $t_1', \ldots, t_n'$ analogously by taking the latest such times. This defines regions $[t_{i-1}', t_i]$ delimited by local minima which are not trivially flat inside. We then let $T_i$ be the earliest time in $[t_{i-1}', t_i]$ such that $E(T_i)$ is a local maximum. Finally, let $s_0 = 0$ and define new endpoints $s_j$ recursively by
\[ s_j = \min\left\lbrace t_i \mid t_i \geq s_{j-1} \: \text{ and } \: E(T_i) - E(t_i) > \sigma \right\rbrace \, , \]

\noindent where $\sigma$ is the standard deviation of $E$. We define $s_j'$ analogously with $t, s$ replaced by $t', s'$ everywhere. This isolates regions $R_j = \left[s_{j-1}',s_j \right]$ which start at the bottom of an energy spike and finish at the other end, ignoring spikes that are too small to be meaningful. The definitions are illustrated in Figure \ref{fig:4.1} below, where one of our BWA spectrograms is overlaid with a depiction of its energy curve.

\vspace{-5pt}
\begin{figure}[h]
\centering
\includegraphics[scale=0.18]{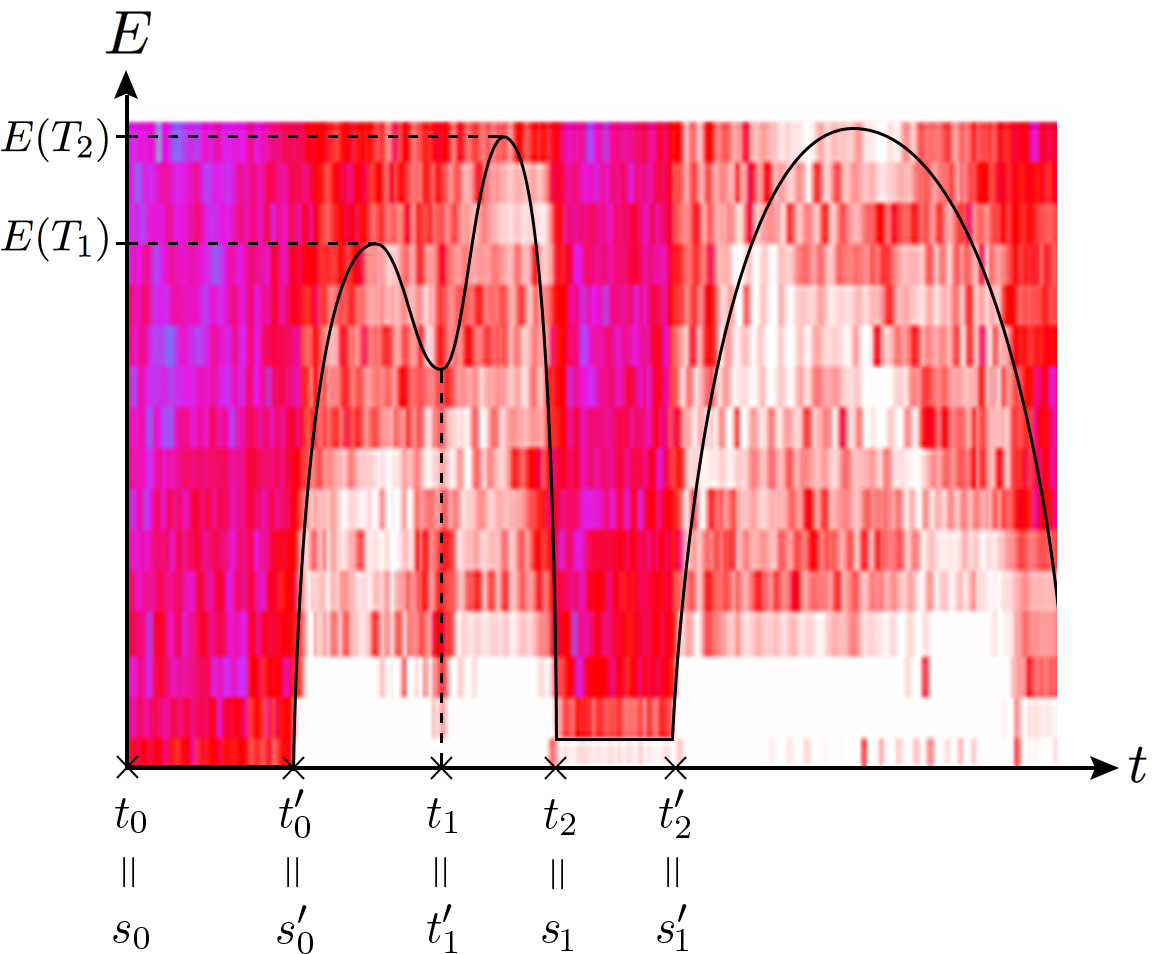}
\caption{Spectrogram overlayed with energy across time.}
\label{fig:4.1}
\end{figure}

\noindent In this example, the local minimum $t_1 = t_1'$ is not equal to any $s_j$ because the energy distance $E(T_1) - E(t_1)$ to the previous maximum is less than $\sigma$. The resulting regions usually contain syllables or short words. In order to form regions of longer words and short phrases, we concatenate these initial regions together. First we choose a cutoff distance $C = 0.02$s and let $k=0$. For each region $R_j=[r_j^1,r_j^2]$, proceed as follows. If
\[ \left|r_{j+k}^2 - r_{j+k+1}^1\right| < C \]
then add a new region $[r_j^1, r_{j+k+1}^2]$, increment $k$, and repeat until the condition is false. Finally, segments shorter than $0.05$s are discarded since any syllable takes longer to pronounce. These contain too little information to be reliably distinguished, and do not provide meaningful repetitions.

\medskip
\noindent
\textbf{Fingerprinting:   }
Following \cite{Haitsma}, our first measure associates a binary rectangular array called \textit{fingerprint} to each region, and computes the percentage of entries at which two arrays differ.
Regions are binned into $N$ non-overlapping windows of length $0.1$s in the time domain, which are then partitioned into $M=32$ bands of equal length between $300$ and $3000$Hz in the frequency domain. We define $E_{n,m}$ to be the energy of window $n$ within frequency band $m$ for $1\leq n \leq N$, $1\leq m\leq M$. We then take second-order finite differences
\[
\Delta^2(n,m) = \left[E_{n,m}-E_{n,m+1}\right]-\left[E_{n-1,m}-E_{n-1,m+1}\right] ,
\]
which provide a discretized measure of curvature in the spectral energy distribution over time.
The value of the fingerprint at position $(n,m)$ is now defined as $F(n,m) = \indfun{\Delta^2(n,m) > 0} \, . $
Given a fingerprint pair, the percentage of positions at which arrays differ provides a measure $E$ of dissimilarity between regions.

\medskip
\noindent
\textbf{Correlation:   }
The second metric, based on \cite{Herley}, makes use of the correlation between Fourier coefficients over short windows.
Regions $R_1$ and $R_2$ are first split into $N$ overlapping windows $w^i_1,\ldots,w^i_N$ for $i=1,2$. For each window $w^i_n$, let $\hat{w}^i_n(m)$ be the Fourier coefficients corresponding to frequencies between $300$ and $3000$Hz. For each $m$, we compute the correlation $C(m) = \frac{\sigma_{1,2}(m)}{\sigma_1(m)\sigma_2(m)}$ between the values of the $m^\mathrm{th}$ coefficient of the two regions, where 
\[ \sigma_i(m) = \sqrt{\sum_{n=1}^N(\hat{w}^i_n(m)-\hat{w}^i(m))^2} \: , \]
\[ \sigma_{1,2}(m) = \sum_{n=1}^N(\hat{w}^1_n(m)-\hat{w}^1(m))(\hat{w}^2_n(m)-\hat{w}^2(m)) \, , \]
and $\hat{w}^i(m) = \frac{1}{N}\sum_{n=1}^N \hat{w}_n^i(m)$. 
\noindent Finally, averaging $C(m)$ over $m=1,\ldots,M$ 
yields an overall similarity measure $C$ for $R_1$ and $R_2$.
This measure is less sensitive and produces more false positives than fingerprints. On the other hand, correlation can pick up on noisy repetitions where fingerprints fail. Our approach is to combine these methods so as to balance their strengths and weaknesses.

\subsection{Scoring}
\label{ssec: scoring}
Combining the fingerprint and correlation metrics into a single score, define $S(E,C) = \sqrt{f_1(E)f_2(C)}$ where
\[ f_1(E) = \indfun{E < 0.3} + \indfun{0.3 \leq E \leq 0.45} \left[ \frac{20}{3}(0.3-E) + 1 \right] \, , \]
\[ f_2(C) = \indfun{C > 0.55} + \indfun{0.25 \leq C \leq 0.55} \left[ \frac{10}{3}(C-0.25) \right] \, . \]
The functions $f_1$  and $f_2 $ are designed to convert the outputs of each method to more meaningful levels of confidence that can be compared and combined, taking into account our empirical observations about the behavior of each method. For example, both our experiments and the paper \cite{Haitsma} suggest that a fingerprint difference above $45\%$ corresponds to regions that are almost certainly not repetitions. Similarly, we are almost certain that a fingerprint difference below $30\%$ corresponds to repeated regions.
After evaluating segments, the measures are aggregated to score the entire audio file. This total score is computed as the average of non-zero scores among the top $5\%$ of unique comparisons. As such, this score is higher for files that contain more or clearer repetitions, and lower for those with fewer or less distinguishable repetitions.

Though repetition tends to be more frequent in scenarios of conflict, significant disputes can further be distinguished from mild ones via a measure of intensity. High conflict scenarios often involve shouting or loud commands, producing higher energy.
Accordingly, an intensity score is computed by averaging the energy among the same top $5\%$ set of repetitions. The overall conflict score for an audio signal is the product of its repetition and intensity scores.

\section{Results and discussion}
\label{sec:results}
\begin{figure}[t]
\begin{center}
	\includegraphics[scale=0.26]{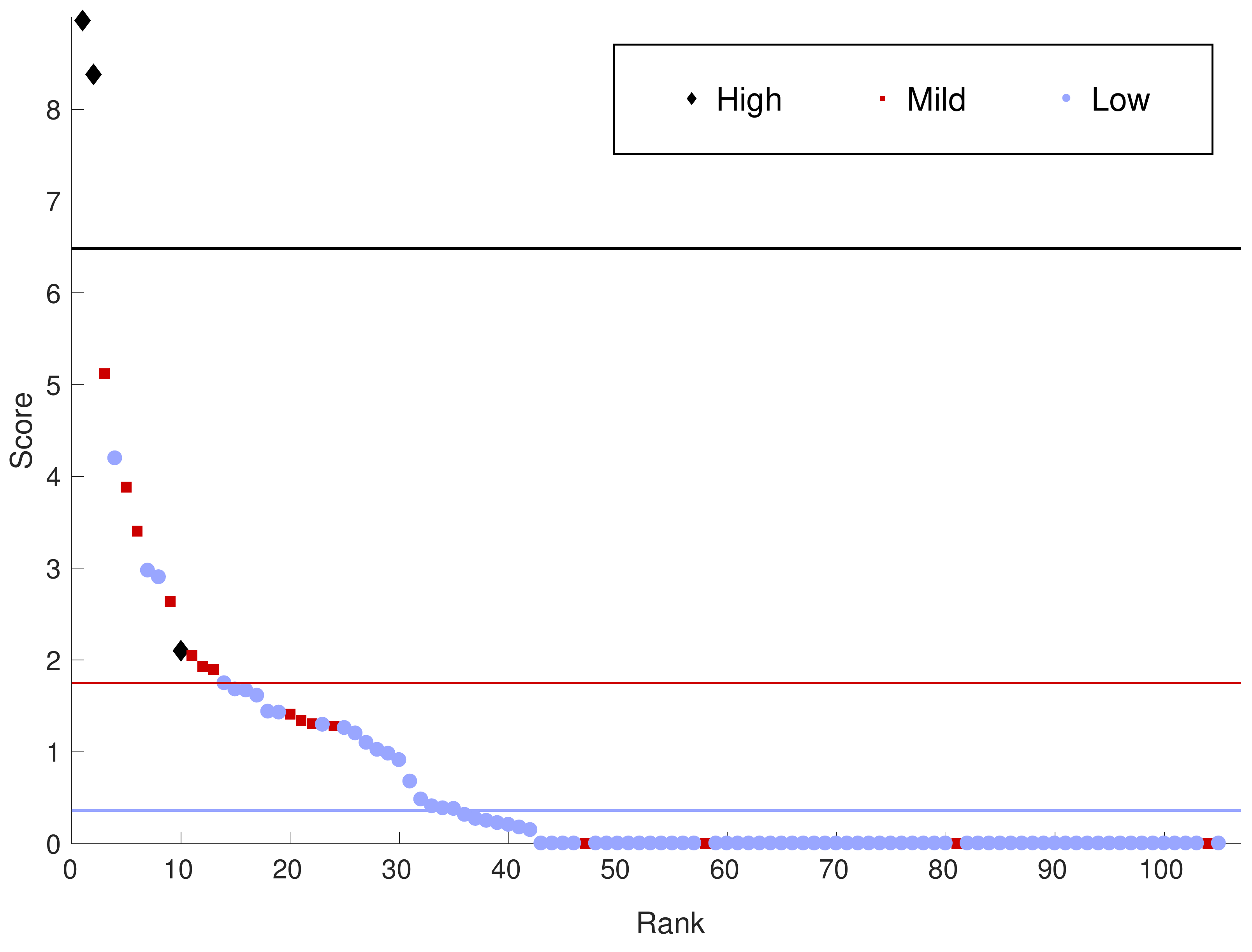}
\end{center}
\vspace{-5pt}
\captionsetup{justification=centering}
\caption{Plot of conflict score against rank. Horizontal lines depict the mean score for the class of corresponding color.}
\label{fig:4.2}
\end{figure}

We test our approach on a collection of $105$ body-worn audio files provided by the LAPD, of lengths between $3$ and $30$ minutes each. The files are manually labeled according to level of conflict, where the classes and criteria are as follows:
\begin{itemize}[itemsep = 0.1pt]
	\item[\textbf{2.}] High conflict (3 files): active resistance, escape, drawing of weapon, combative arguments.
	\item[\textbf{1.}] Mild conflict (15 files): questioning of officer judgment, avoiding questions, avoiding to comply with commands, aggressive tone.
	\item[\textbf{0.}] Low conflict (87 files): none of the above.
\end{itemize}
Figure \ref{fig:4.2} is a plot of files ranked in descending order of conflict score as determined by our method, illustrating that those labeled as high or mild conflict are concentrated toward the top. More specifically, all three files labeled as high conflict occur in the top $10$ scores.

In general, the three classes are correctly prioritized by the scoring algorithm. Only $4$ of the $18$ files in classes $1$ and $2$ fell below rank $24$. In other words, $78\%$ of the files with any conflict would be found by reviewing only the top $23\%$ of files in the list. The mean scores for each class, displayed in the figure, are clearly well-separated.
Our method can thus be used to significantly reduce the time it takes to manually locate files containing conflict. Further, the algorithm automatically isolates the repetitions detected in a given file, which amount to very short audio portions relative to the entire signal. As such, we may quickly search through the high-rank audio files by listening to these portions.

Given a larger dataset, one could automatically determine the adequate scores to label a file as containing high/mild/low conflict using a learning algorithm of choice. One may also input the fingerprinting, auto-correlation and intensity measures as features into the learning algorithm, producing a decision hyperplane in three dimensions.

In addition to their immediate use, our findings may also inform policy to better aid future work.
We find that officer speech is vastly more informative than other voices, which are less comprehensible and contribute to false positives. To further improve performance, one may exclude all speech except that of the officer.
This falls under the task of speaker diarization---see \cite{Anguera} for a recent review---and most studies in this area are based on relatively clean data (broadcast meetings, conference calls). State-of-the-art methods including \cite{Friedland} and \cite{Fox} achieve no less than $18\%$ diarization error rate on average, rising to $30\%$ for some of the meetings, but perform much worse when applied to our BWA data. This obstacle may be overcome provided additional labeled data. Given a sample of the officer's voice that can be used to identify them elsewhere, our supervised learning task translates to speaker verification \cite{Reynolds}. Such data could be provided by requiring officers to record a few minutes of clean speech once in their career; this sample could then be overlaid with non-speech extracted by our pipeline to render it comparable with BWA files featuring a range of noise environments.

\section{Conclusion}
\label{sec:conclusion}
To summarize, we offer a novel method for automatic conflict detection which is successful for applications in police body-worn audio. We are able to automatically select audio files which are very likely to contain conflict, despite a small number of high conflict files. We propose eliminating non-officer speech through speaker verification and using all three sub-scores as learning features to improve these results.

\vfill\pagebreak

\bibliographystyle{IEEEbib}
\bibliography{refs}

\end{document}